\documentclass[reprint, amsmath,amssymb,aps]{revtex4-1}

\usepackage[english]{babel}
\usepackage{xcolor}
\usepackage{url}
\usepackage{graphicx}
\usepackage{dcolumn}
\usepackage{bm}

\newcommand{\Ham}{H}

\usepackage{hyperref}
\hypersetup{
    colorlinks=true,
    linkcolor=blue,
    filecolor=magenta,      
    urlcolor=cyan,
} 
\begin{document}


\title{Self-learning Monte Carlo with equivariant Transformer}

\author{Yuki Nagai}
\email{nagai.yuki@mail.u-tokyo.ac.jp}
\affiliation{Information Technology Center, The University of Tokyo, 6-2-3 Kashiwanoha, Kashiwa, Chiba 277-0882, Japan}
\affiliation{CCSE, Japan Atomic Energy Agency, 178-4-4, Wakashiba, Kashiwa, Chiba 277-0871, Japan}
\author{Akio Tomiya}%
\email{akio@yukawa.kyoto-u.ac.jp}
\affiliation{
Department of Information and Mathematical Sciences, Tokyo Woman’s Christian University, Tokyo 167-8585, Japan
}
\affiliation{%
Faculty of Technology and Science, International Professional University of Technology, 3-3-1, Umeda, Kita-ku, Osaka, 530-0001, Osaka, Japan
}%

\date{\today}

\begin{abstract}
Machine learning and deep learning have revolutionized computational physics, particularly the simulation of complex systems. Considering equivariance and long-range correlations is essential for simulating physical systems. Equivariance imposes a strong inductive bias on the probability distribution described by a machine learning model, while long-range correlation is important for understanding classical/quantum phase transitions.
Inspired by Transformers used in large language models, which can treat long-range dependencies in the networks, we introduce a symmetry equivariant Transformer for self-learning Monte Carlo. We evaluate our architecture on a spin-fermion model (i.e., the double exchange model) on a two-dimensional lattice. Our results show that the proposed method overcomes the poor acceptance rates of linear models and exhibits a similar scaling law to large language models, with model quality monotonically increasing with the number of layers. Our work paves the way for the development of more accurate and efficient Monte Carlo algorithms with machine learning for simulating complex physical systems.

\end{abstract}

\maketitle

\section{Introduction}
The fields of machine learning and deep learning are rapidly transforming computational physics, with profound implications for fields such as condensed matter, particle physics, and cosmology. 
This progress has been particularly notable since 2016 \cite{silver2016mastering}, with the emergence of innovative techniques catalyzed by the rise of deep learning.

One of the most promising applications of machine learning in computational physics is the use of trainable models in Monte Carlo methods. 
The Self-learning Monte Carlo (SLMC) method, developed for electron systems in condensed matter \cite{Liu_2017}, has been widely used in strongly correlated electron systems \cite{Liu2017-wl,Shen2018-fl,Kohshiro2021-ea,Nagai2017-ux}, molecular dynamics \cite{Nagai2020-gx,Nagai2020-mf,Kobayashi2021-lf,Nagai2024-ff}, and lattice quantum chromodynamics (QCD) \cite{Nagai2023-ui,Nagai2021-pw}. 
SLMC constructs effective models to generate the Boltzmann weight by means of machine learning from gathered original Markov chains. 
Moreover, it has been used in flow-based sampling algorithms to realize field theoretical distributions using trivial distribution and Jacobian \cite{Albergo_2019,Kanwar_2020,Boyda_2021,Albergo_2021,hackett2021flowbased,Albergo_2022,Abbott_2022,abbott2022sampling,abbott2023normalizing,Tomiya:2022meu}.

The Transformer architecture, which debuted as a translation model to consider long-range dependencies in sentences in natural language \cite{vaswani2017attention}, has since been applied to a wide range of problems, including image recognition \cite{dosovitskiy2021image}, protein folding computations \cite{Jumper2021}, and foundation models such as GPT (Generative Pre-trained Transformer) \cite{radford2018improving, radford2019language, brown2020language, openai2023gpt4}.
Transformer models exhibit a favorable scaling law for model capacity, meaning that their performance increases in proportion to the size of the input data \cite{kaplan2020scaling, lin2021survey}.

The symmetry of a system is essential to its description in physics. 
In machine learning, the imposition of an inductive bias in probability models corresponds to the symmetry of the system \cite{cao2021random, koyama2023neural}. 
For example, convolutional layers for neural nets respect translational symmetry, which helps in image recognition.
This concept has been generalized to general symmetries \cite{cohen2016group}, and is applied to a large class of models in physics \cite{Albergo_2019,Kanwar_2020,Boyda_2021,Albergo_2021,hackett2021flowbased,Albergo_2022,Abbott_2022,abbott2022sampling,abbott2023normalizing}.

Machine learning-assisted Monte Carlo methods combined with effective theory have achieved notable success in many areas \cite{Liu_2017,Liu2017-wl,Shen2018-fl,Kohshiro2021-ea,Nagai2017-ux,Nagai2020-gx,Nagai2020-mf,Kobayashi2021-lf,Nagai2023-ui,Nagai2021-pw,Albergo_2019,Kanwar_2020,Boyda_2021,Albergo_2021,hackett2021flowbased,Albergo_2022,Abbott_2022,abbott2022sampling,abbott2023normalizing,Tomiya:2022meu}.
However, capturing the global correlations arising from fermionic degrees of freedom, which are inherent to quantum systems, remains a significant challenge. Traditional machine learning methods, which are often designed for classical data and its convolutions, are ill-suited to capture the unique correlations present in quantum systems. To effectively address this challenge, we need to introduce a method that can grapple with both the quantum global correlations and the symmetries inherent to a physical system.

In this study, we develop a Transformer for physical systems with fermions and global symmetries.
As a example, we apply Transformer on a two-dimensional spin-fermion model ({\it i.e.}, double exchange model) within SLMC.
Leveraging the extensive capacity of the attention layer, we successfully construct an effective model of the system using the equivariant Transformer. We find that the model with the attention blocks overcomes the poor acceptance rates of linear models and exhibits a similar scaling law as in large language models.


\section{Model}
\subsection{Double exchange model}
In this paper, we focus on the system with fermions and classical spins on two-dimensional square lattice: 
\begin{align}
    \Ham = \Ham_{\rm F} + \Ham_{\rm ex},
\end{align}
where 
\begin{align}
    \Ham_{\rm F} &\equiv -t \sum_{\alpha,\langle i,j \rangle} (\hat{c}^{\dagger}_{i\alpha} \hat{c}_{j\alpha} + {\rm h.c.})  - \mu \sum_{\alpha,i} \hat{c}_{i\alpha}^{\dagger} \hat{c}_{j\alpha}, \\
    \Ham_{\rm ex} &\equiv \frac{J}{2} \sum_{i} {\bm S}_i \cdot \hat{\bm \sigma}_i.
\end{align} 
Here ${\bm S}_i$ is the classical Heisenberg spins on the $i$-th site, $t$ is the hopping constant, $\hat{c}_{i\alpha}^{\dagger}$ ($\hat{c}_{i\alpha}$) is the fermionic creation (annihilation) operator at the $i$-th site for fermion with spin $\alpha \in \{ \uparrow,\downarrow\}$, $\langle i,j \rangle$ are the pairs of nearest neighbors, $J$ is the interaction strength between the classical spins and the electrons, and the Pauli matrices are defined as $[\hat{\bm \sigma}_i]_{\gamma} \equiv \hat{c}_{i\alpha}^{\dagger} \sigma_{\alpha \beta}^{\gamma} \hat{c}_{i\beta}$ ($\gamma = x,y,z$).
We consider the hopping constant $t$ as the unit of energy. 
We adopt the periodic boundary condition on $N_x \times N_y$ site system. The total number of the sites is $N \equiv N_x N_y$.
This model is called double exchange (DE) model with classical spins \cite{Barros2013-op,Liu2017-wl,Stratis2022-zr}.
The Hamiltonian has $O(3)$ rotational symmetry in the spin sector and discrete translational invariance. 

\subsection{Long range correlation}
An effective theory can have non-local long range interactions even if an original Hamiltonian has short range interactions. This happens in mixed systems with fermions and others. 
For example, the spin model with long-range spin-spin interaction so-called Ruderman–Kittel–Kasuya–Yosida (RKKY) interaction can be regarded as the effective model of the DE model in the limit of the weak coupling \cite{Ruderman1954-ox,Kasuya1956-el,Yosida1957-ui}.
Models with RKKY interaction has been used to describe spin-glass order in the diluted magnetic alloys \cite{Mydosh}. 

\section{Effective models}
\subsection{General effective model}
Before introducing the effective model with the Transformer architecture, let us consider general effective model that is an extension of the linear model. 
The general effective model should be invariant with respect to rotations in spin space, spatial reflections, and translations in real space. 
By introducing the spin rotation matrix $R$ which rotates all spins in spin space, the effective model should satisfy the following relation: 
\begin{align}
    \Ham = f( {\bm S} R  ) = f( {\bm S}  ).
\end{align}

In this work, we regard the classical spin field as the matrix whose matrix element is defined as $S_{i \mu} \equiv [{\bm S}_i]_{\mu}$.
One of the appropriate functions with the rotational invariance is the functional of the Gram-matrix \cite{Xu2021-tv}: 
\begin{align}
\Ham = f_G(G( {\bm S})).
\end{align}
where  the Gram matrix is rotation-invariant, the size of the Gram matrix is the number of sites $N$ defined as $G(  {\bm S}) \equiv {\bm S} {\bm S}^T$. 
According to Ref. \cite{Xu2021-tv}, if the Gram matrix of two spin field are equal, their geometrical structures are also the same.

To extend the linear model, we introduce the effective spin field ${\bm S}_{\rm eff}$ defined as 
\begin{align}
    {\bm S}_{\rm eff} = \check{M}(G( {\bm S})) {\bm S},
\end{align}
where $\check{M}(G( {\bm S}))$ is a $N \times N$ matrix and  ${\bm S}_{\rm eff}$ is equivarient with respect to spin rotations since the Gram matrix is rotational invariant.
With the use of this effective spin field, we propose the effective model expressed as 
\begin{align}
\Ham_{\rm eff}^{\rm NN} = {\rm Tr} \left[  {\bm S}_{\rm eff} (\hat{J} {\bm S}_{\rm eff})^{\top} \right] + E_0, \label{eq:heff}
\end{align}
where $\check{J}$ is a $N \times N$ matrix.
This model is a functional of the Gram matrix with respect to ${\bm S}_{\rm eff} $.
Here, we introduce the local operator for spins: 
\begin{align}
    \hat{A} {\bm S} &\equiv {\bm S}^{\hat{A}},
\end{align}
where the matrix element of ${\bm S}^{\hat{A}}$ is defined as 
\begin{align}
S_{i \mu}^{\hat{A}} &\equiv  \sum_{\langle i,j \rangle_n} A_n  S_{j \mu}.
\end{align}
$\langle i,j \rangle_n$ indicates $n$-th nearest neighbor.
Since the local operation does not rotate spins in spin space, $\hat{A}$ is an equivariant with respect to rotations in spin space: 
\begin{align}
     \hat{A} ({\bm S} R) = {\bm S}^{\hat{A}} R.
\end{align}
Equivariant networks have been used in various kinds of fields such as point cloud processing \cite{Assaad2022-bi,Deng2021-xg} and machine-learning molecular dynamics \cite{Tholke2022-bg,Batzner2022-ns}.

\subsection{Transformer and attention layer.}
In this paper, we construct the effective spin field that consists of the neural networks with multiple attention layers. 
In this work, we note index of layers as  $l=1,2,\cdots, L$.

\begin{figure}[t]
\begin{center}
\includegraphics[scale=0.35]{ 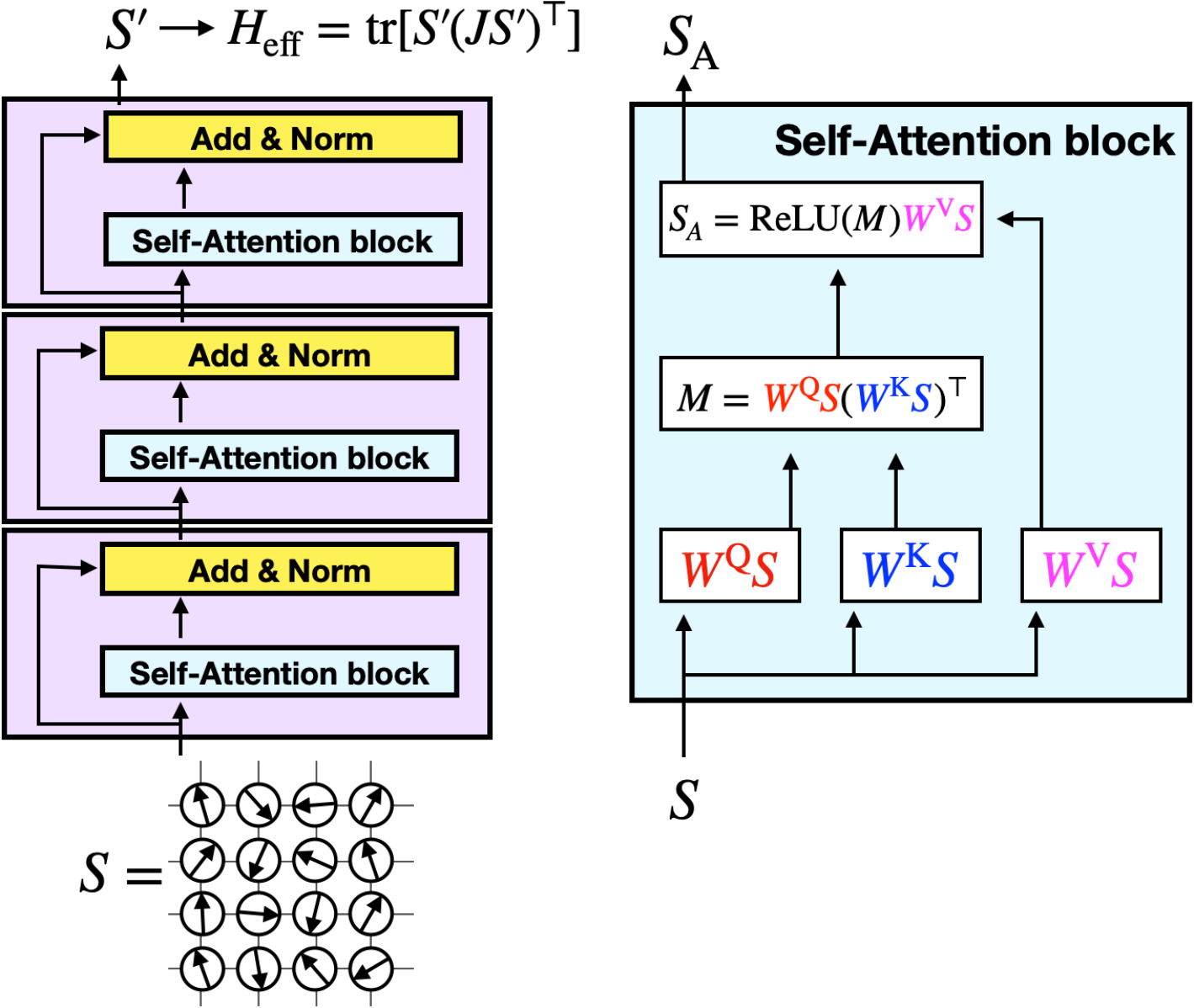}
\end{center}
  \caption{
(\textit{Left})
  Construction of effective Hamiltonian with the use of the equivariant Transformer with three of attention layers. 
  Yellow blocks are defined by Eq. \eqref{eq:normalization_in_transformer}.
  We call purple blocks the attention layers.
(\textit{Right})
Blue blocks are equivariant attention block (See main text).
   \label{fig:attention}}
\end{figure}

The network architecture is expressed by
\begin{align}
    {\bm S}^{(0)} &\equiv  {\bm S}, \\
    {\bm S}^{(l)} &\equiv {\cal N}\left({\bm S}^{(l-1)} + {\rm SelfAttention}_{{\bm \theta}^{(l)}}^{\rm spin}({\bm S}^{(l-1)}) \right), \label{eq:normalization_in_transformer} \\
    {\bm S}_{\rm eff} &=  {\bm S}^{(L)}. \label{eq:transformer}
\end{align}
$ {\rm SelfAttention}_{{\bm \theta}^{(l)}}^{\rm spin}({\bm S}^{(l-1)})$ is a neural network map between spin configurations, which will be explained following.
${\bm \theta}^{(l)}$ represents a set of network trainable parameters in $l$-th layer. 
${\cal N}({\bm S})$  normalizes the spin vector on each lattice site,
\begin{align}
{\cal N}({\bm S}_i) = \frac{{\bm S}_i}{\|{\bm S}_i\|}.
\end{align}
We call this network architecture the equivariant Transformer, which is schematically visualized in Fig.~\ref{fig:attention}.

The attention block is an essential component of the Transformer neural networks (See, Appendix \ref{sec:attention}).  
In our system with classical spin field, we introduce the following self-attention block: 
\begin{align}
    {\rm SelfAttention}^{\rm spin}({\bm S}) = \check{M}{\bm S}^{\rm V},
\label{eq:self_attention}
\end{align}
where
\begin{align}
\left[\check{M} \right]_{ij} = \sigma \left(  \frac{1}{\sqrt{3}} \sum_{\mu} S_{i\mu}^{\rm Q} S_{j\mu}^{\rm K} \right),
\end{align}
and $i,j$ are spatial index. 
$\sigma$ is a nonlinear activation function applied element-wisely.
In this paper, we use the ReLU function. 
Queries, keys, and values are $N \times 3$ matrices defined as 
\begin{align}
{\bm S}^{\rm Q} \equiv \hat{W}^{\rm Q} {\bm S}, \:
{\bm S}^{\rm K} \equiv \hat{W}^{\rm K} {\bm S}, \: 
{\bm S}^{\rm V} \equiv \hat{W}^{\rm V} {\bm S}.
\end{align}
We note that the matrix $\check{M}$ is a functional of the Gram matrix with respect to ${\bm S}$. 
The local operators $\hat{W}^{\rm Q},\hat{W}^{\rm K},\hat{W}^{\rm V}$ have trainable parameters. 
It should be noted that the number of the trainable parameters in the local operator is usually fewer than a dozen. 
For example, if one considers $n$-th neighbors, the number of the trainable parameters in one local operator is only $n+1$. 
In terms of machine learning, this is called the weight sharing. 


\subsection{Equivariance}
We discuss a concept of equivariance. 
A function $f: X \rightarrow Y$ is equivariant with respect to a group $G$ that acts on $X$ and $Y$ if:
\begin{align}
D_{Y}[g]f(x) = f(D_{X}[g]x),  \: \: \forall g \in G, \forall x \in X,
\end{align}
where $D_X[g]$ and $D_Y[g]$ are the representations of the group element $g$ in the spaces $X$ and $Y$, respectively. 

The effective spin field generated by our attention layer has the spin rotational equivariance. 
We show that the effective spin has also the translational operation equivariance as follows. 
Since the matrices $\hat{W}^{\rm Q}$, $\hat{W}^{\rm K}$ and $\hat{W}^{\rm V} $ are local operators, the spin fields ${\bm S}^{\rm Q}$, ${\bm S}^{\rm K}$ and ${\bm S}^{\rm V}$ are translational operation equivariant ($T {\bm S}^{\rm Q} =  \hat{W}^{\rm Q} (T{\bm S})$). 
Here, $T$ is a translational operator. 
Although the matrix $\check{M}$  is not invariant with respect to the translational operation ($\check{M} \rightarrow T \check{M} T^{\top}$), the self-attention block is translational operation equivariant because $\check{M} {\bm S}^{\rm V} \rightarrow T \check{M} T^{\top} ({T \bm S}^{\rm V}) = T (\check{M} {\bm S}^{\rm V})$.

\section{Demonstration}
We show that the effective model with the Transformer architecture can capture the long-range correlation in the original model. 
The partition function of the DE model is expressed as 
\begin{align}
{\cal Z} = \sum_{\{ {\bm S} \} } W(\{ {\bm S} \} ),
\end{align}    
where $W(\{ {\bm S} \} ) \equiv \prod_{n} (1+e^{-\beta(\mu - E_n(\{ {\bm S} \}  ))})$ is the Boltzmann weight with as a spin configuration ${\{\bm S} \}$ and $E_n(\{{\bm S} \})$ is the $n$-th eigenvalue of the Hamiltonian. 
A standard method to simulate this system is the Markov chain Monte Carlo (MCMC) method. 
In the DE model, however, computational complexity is huge due to a fermion determinant for itinerant electrons, since a full diagonalization is needed to estimate the fermion determinant for given spin configurations for each Monte Carlo (MC) step. 
Many alternative algorithms to simulate electrons coupled to classical spin fields have been developed \cite{ALONSO2001587,FURUKAWA2001410,doi:10.1143/JPSJ.73.1482,ALVAREZ200532,Alvarez_2007,Barros2013-op,Liu2017-wl,Kohshiro2021-ea}.
\subsection{Self-learning Monte Carlo}

We use the SLMC, one of the Markov chain Monte Carlo method with effective models (See, Appendix \ref{sec:slmc}). 
In the SLMC, we introduce an effective model for making a proposal in the Markov chain with the probability $W_{\rm eff}(\{ {\bm S} \}) \equiv W_{\rm prop}(\{ {\bm S} \})$. 
The probability $W_{\rm eff}(\{ {\bm S} \})$ can be written as 
\begin{align}
W_{\rm eff}(\{ {\bm S} \}) = \exp \left[ -\beta \Ham_{\rm eff}(\{{\bm S} \},T) \right],
\end{align}
where the effective Hamiltonian $\Ham_{\rm eff}(\{{\bm S} \},T)$ is a real scalar. 
The acceptance ratio in the SLMC is given as 
\begin{align}
    A( \{ {\bm S}' \}, \{ {\bm S} \}) = {\rm min} \left( 1, \frac{W(\{ {\bm S}' \} ) }{W(\{ {\bm S} \} ) } 
    \frac{W_{\rm eff}(\{ {\bm S} \} ) }{W_{\rm eff}(\{ {\bm S}' \} ) } 
\right). 
\end{align}
If we can design the proposal Markov chain whose probability is equal to that of the original Markov chain, the proposed spin configuration $\{ {\bm S}' \}$ is always accepted. 
We note that  the effective Hamiltonian generally depends on the temperature. 
The average acceptance ratio $\langle A \rangle$ can be estimated by \cite{Shen2018-fl} 
\begin{align}
    \langle A \rangle = e^{-\sqrt{{\rm MSE}}}, \label{eq:mse}
\end{align}
with the mean squared error (MSE) between the original and effective models: 
\begin{align}
    {\rm MSE} = \frac{1}{N_{\{{\bm S} \} }} \sum_{\{{\bm S} \}} \left|\log W(\{{\bm S} \})-\log W_{\rm eff}(\{{\bm S} \})
    \right|^2,
\end{align}
where $N_{\{{\bm S} \} }$ is the number of spin configurations.

\subsection{Model training}
The trainable weights in the $l$-th layer are $ \hat{W}^{{\rm Q},(l)}$, $ \hat{W}^{{\rm K},(l)}$ and $ \hat{W}^{{\rm V},(l)}$. 
The number of the trainable parameters in the local operator $\hat{W}^{\alpha}$ is $m^{\rm \alpha}$.
In this paper, we consider $n$-th nearest neighbors on all attention layers so that $m^{\rm \alpha} = n+1$.
There are $3(n+1)$ trainable parameters in each attention layer. 
As shown in Eq.~(\ref{eq:heff}), the trainable weights in the last layer are $\hat{J}$ and $E_0$. 
The number of the trainable parameters in $\hat{J}$ is $m_{J}+1$ since $m_{J}$-th nearest neighbors are considered in the last layer. 
The total number of the trainable parameters with $L$ attention layers is $3L(n+1) + m_J + 1 + 1$, which is quite less than fully-connected neural networks. 
%
We remark that if the trainable weights in $l$-th layer are zero,  ${\bm S}^{{\rm eff}(L)} $ falls back to ${\bm S}^{{\rm eff}(L-1)} $.
Because the parameter space for ${\bm S}^{{\rm eff}(L-1)}$ is inside the parameter space for   ${\bm S}^{{\rm eff}(L)}$, we can use converged parameters for ${\bm S}^{{\rm eff}(L-1)} $ as the initial guess for ${\bm S}^{{\rm eff}(L)} $.

We should note that the number of the trainable parameters does not depend on the system size $N$. 
Therefore, the effective model obtained in a smaller system can be used in larger systems without any change. 

We train the effective model with the use of the SLMC. 
We set the total number of the Metropolis-Hastings tests with the original model $N_{\rm MC}^{\rm original}$ and the length of the proposal Markov chain $N_{\rm MC}^{\rm eff}$, 
where ${\bm S}^{{\rm eff}(L-1)}$ is the effective spin field with $L-1$ attention layers. 
The training procedure with the use of the SLMC is shown as follows. 
At first, we prepare the initial effective model. 
The initial guess is produced by the iterative method proposed in the previous section. 
Next, we produce a randomly oriented spin configuration. 
With the use of the MCMC with the effective model $\Ham_{\rm eff}$, the random spin configurations are approximately thermalized. 
This configuration is regarded as the initial configuration $\{ {\bm S} \}_1$ in terms of the SLMC.  
Then, we calculate the weights $W$ and $W_{\rm eff}$. 
The proposal configuration $\{ {\bm S} \}_2$ is generated by the MCMC with the effective model.  
The configuration $\{ {\bm S} \}_2$ is accepted if the uniform random number is smaller than the acceptance ratio $A(\{ {\bm S}_2\}, \{ {\bm S}_1 \} )$. 
If the configuration is rejected, we use  $\{ {\bm S} \}_1$ as the initial configuration for the effective MCMC and try to propose new configuration $\{ {\bm S} \}_2'$. 
The quality of the effective model can be estimated by the average acceptance ratio in the SLMC. 
If the acceptance ratio in the SLMC is too low, we make  $N_{\rm MC}^{\rm eff}$ smaller. 
In this paper, we first use  $N_{\rm MC}^{\rm eff} = 10$ to obtain better initial guess and set  $N_{\rm MC}^{\rm eff} = 100$ in the main SLMC.

\subsection{Iterative training}

At first, we consider the linear model ($L = 0$) shown in Eq.~(\ref{eq:heff}) where the trainable weights are only $\hat{J}$ and $E_0$. 
After the training with the linear model, we introduce the weights $ \hat{W}^{{\rm Q},(l=1)}$, $ \hat{W}^{{\rm K},(l=1)}$ and $ \hat{W}^{{\rm V},(l=1)}$ whose matrix elements are uniform random numbers $r \in [-\epsilon,\epsilon]$. 
In this paper, we set $\epsilon = 10^{-6}$.
After the training with the effective model with $l$ attention layers, we can prepare the initial guess for the effective model with $l+1$ attention layers. 
To optimize the training parameters, we use \texttt{Flux.jl} \cite{innes2018fashionable}, machine-learning framework written in Julia language \cite{bezanson2015julia}. 
We adopt the AdamW optimizer (the parameters are $\eta = 0.001$ and $\beta = (0.9,0.999)$ as in \cite{loshchilov2019decoupled}) and the size of the minibatch is 100, which means that the effective model is improved at every 100 Metropolis-Hastings tests.


\section{Results}

\subsection{Parameter setup}
We consider two-dimensional $N_{x} \times N_y$ square lattice. 
The interaction strength $J$ between the classical spins and the electrons and the chemical potential $\mu$ are  set to $J = 1t$ and $\mu = 0$, respectively. 
The effective model is trained by the SLMC in the $6 \times 6$ square lattice. 
We consider $6$-th nearest neighbors in the attention layers so that we set $m^{\rm \alpha} = n+1 = 7$.
In the each proposal Markov chain, we adopt the local spin rotation update. 

For comparison, we also perform the MCMC simulation with the original Hamiltonian. 
In the $6 \times 6$ square lattice system, the length of the Markov chain is $2 \times 10^5$.

\subsection{Physical observable calculated by the SLMC}
Since the SLMC is the MCMC with the original Hamiltonian, the physical quantities can be calculated exactly. 
Over the whole temperature range, we use same effective model trained at $T = 0.05t$. 

\begin{figure}[t]
\begin{center}
\includegraphics[width=\linewidth]{ 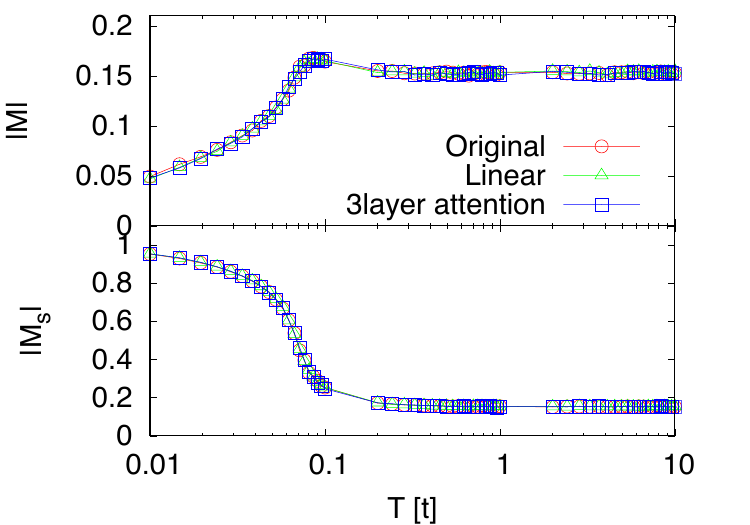}
\end{center}
  \caption{Magnitude of average magnetization and staggered magnetization for a two-dimensional system with $6 \times 6 = 36$ lattice sites. 
For each temperature, we generate $2 \times 10^5$ samples using exact diagonalization (red circles), $5000$ samples using SLMC with the linear model (green triangles) and the effective model with 3layer attention (blue squares). 
   \label{fig:SLMCresults}}
\end{figure}

As shown in Fig.~\ref{fig:SLMCresults}, the SLMC with effective models successfully reproduces the physical quantities obtained by the original model~\footnote{We define magnitude of average magnetization 
   $ |{\bm M}| = \frac{1}{Z}  \sum_{\{ {\bm S} \} }  W(\{ {\bm S} \} ) \left\|\frac{1}{N} \sum_{i} {\bm S}_i \right\|$ and staggered magnetization $
     |{\bm M}_s| = \frac{1}{Z}  \sum_{\{ {\bm S} \} }  W(\{ {\bm S} \} ) \left\|\frac{1}{N} \sum_{i} (-1)^{\sum_{j=1}^d x_j }{\bm S}_i\right\|$, 
where $x_j$ in the sum $\sum_{j}^d x_j$ indicates the site index in the $j$-th dimension. }. 
There is the anti-ferromagnetic order in low temperature regime~\cite{Stratis2022-zr}.

\subsection{Autocorrelation time}
\begin{figure}[th]
\begin{center}
\includegraphics[width=0.9 \linewidth]{ 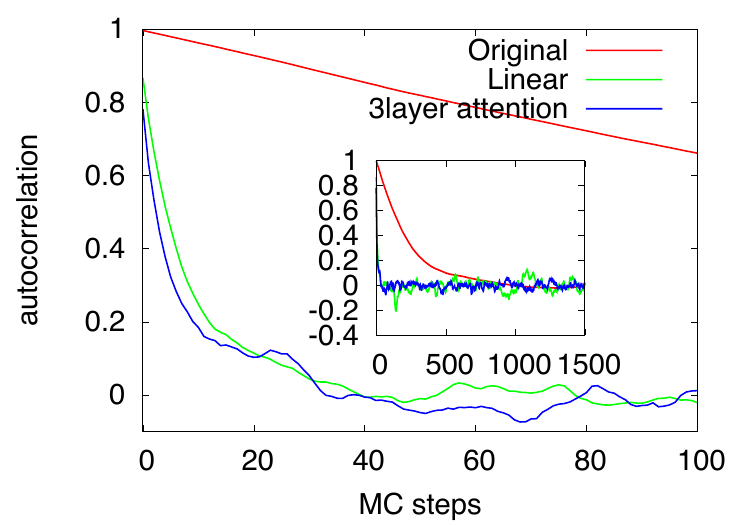}
\end{center}
  \caption{(Color online) Autocorrelation for magnitude of the staggered magnetization for a two-dimensional system with $6 \times 6 = 36$ lattice sites at $T = 0.01t$. 
   \label{fig:SLMCautocorrelation}}
\end{figure}

We confirm that the autocorrelation time in the SLMC is drastically shorter than that in the original MCMC. 
We show the autocorrelation at the lowest temperature ($T = 0.01t$) in Fig.~\ref{fig:SLMCautocorrelation}.
The SLMC reduces the autocorrelation where the number of the MC steps means the number of the calculations of the original model \cite{Liu2017-wl,Nagai2017-ux,Nagai2020-mf,Kohshiro2021-ea}. 

\subsection{Long-range interaction in the attention layers}
\begin{figure}[th]
\begin{center}
\includegraphics[width=\linewidth]{ 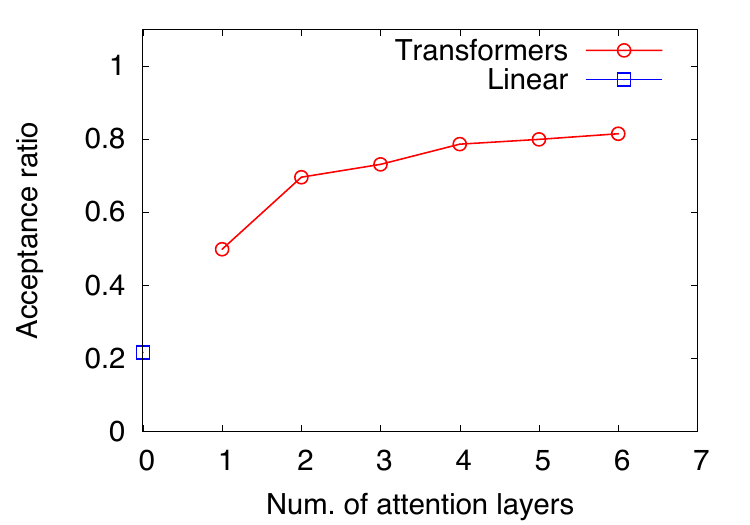}
\end{center}
  \caption{Acceptance ratio in the SLMC with the effective model. We only consider the nearest neighbors in the last layer ($m_J = 1$). A blue square indicate the acceptance ratio for the linear model. Red circles indicate models with attention blocks, with $L = 1, 2, 3, 4, 5, 6$ from the left. \label{fig:layerdep}
   }
\end{figure}

Next, we show that our model with attention layers can capture the long-range interaction. 
In this section, we consider the nearest neighbors in the last layer ($m_J = 1$). 
Since a quality of the effective model can be estimated by the average acceptance ratio of the SLMC, we consider the layer number dependence of the acceptance ratio. 
The effective models are trained at $T = 0.05t$ on the $6 \times 6$ square lattice. 
We set the total number of the Metropolis-Hastings tests with the original model $N_{\rm MC}^{\rm original} = 3 \times 10^4$  and the length of the proposal Markov chain $N_{\rm MC}^{\rm eff} = 100$ in this section.  
The average acceptance ratio for the SLMC with the linear model is only about 21\%, since the long-range spin-spin interaction is neglected in this model. 
The layer number dependence is shown in Fig.~\ref{fig:layerdep}.
The acceptance ratio becomes higher with increasing the number of attention layers.

\subsection{Physical observable calculated by the effective model}
\begin{figure}[t]
\begin{center}
\includegraphics[width=0.9 \linewidth]{ 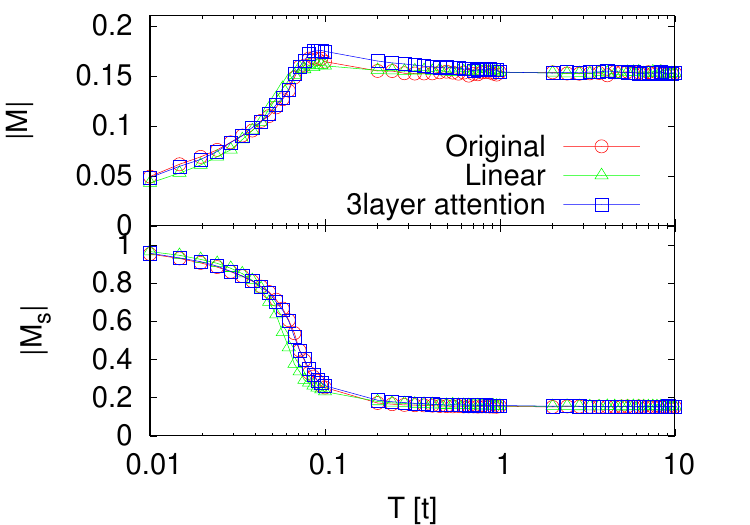}
\end{center}
  \caption{(Color online) Magnitude of average magnetization and staggered magnetization for a two-dimensional system with $6 \times 6 = 36$ lattice sites without using the SLMC. 
For each temperature, we generate $2 \times 10^5$ samples using exact diagonalization (red circles), the linear model (green triangles) and the effective model with 3layer attention (blue squares). 
   \label{fig:effresults}}
\end{figure}

In the SLMC, one has to calculate the Boltzmann weights of the original Hamiltonian. 
However, if the effective model is similar to the original model, the MCMC only with the effective model can be used as the "original" MCMC. 
As shown in Fig.~\ref{fig:effresults}, the MC with effective models almost reproduces the physical quantities obtained by the original model. 
In the case of the linear model, the staggered magnetization slightly differs from that with the original model in low temperature region. 
The effective model constructed using the transformer architecture has sufficient capability to capture the original model. 

We should note that, since the derivative of the effective model can be easily calculated by the back-propagation technique, one can use the Hybrid Monte Carlo method \cite{ALONSO2001587} or the Langevin Dynamics \cite{Barros2013-op} with the effective model.

\section{Discussion}

\subsection{Scaling law}
\begin{figure}[t]
\begin{center}
\includegraphics[width=\linewidth]{ 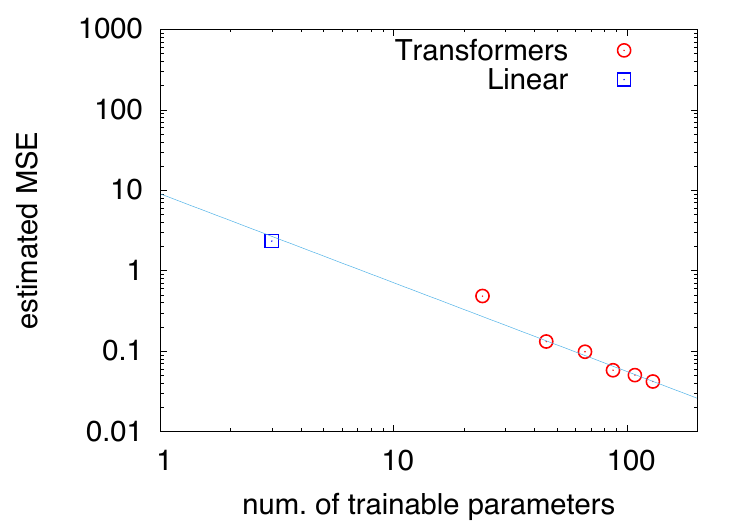}
\end{center}
  \caption{Estimated MSE as a functional of the number of trainable parameters. 
  We only consider the nearest neighbors in the last layer ($m_J = 1$).
  Blue square corresponds to the estimated MSE for the linear model.
  Red circles indicate different models with $L=1,2,3,4,5,6$ from the left.
  Only points for $L\geq2$ are fitted and the point for the linear model is not included.
\label{fig:scaling}
   }
\end{figure}

It is known that there is the empirical scaling laws for language model performance \cite{kaplan2020scaling}. 
The authors in Ref. \cite{kaplan2020scaling} claims that performance of language models improves smoothly as we increase the model size, dataset size, and amount of compute used for training and that empirical performance has a power-law relationship with each individual factor when not bottlenecked by the other two. 
We show that our model with attention blocks has also a similar scaling law.
In the previous section, we discuss the layer-number dependence of the acceptance ratio (Fig. \ref{fig:layerdep}). 
With the use of Eq.~(\ref{eq:mse}), the MSE can be estimated from the average acceptance ratio: 
\begin{align}
{\rm MSE} = 
\big(
\log \langle A \rangle
\big)^2.
\end{align}
Figure \ref{fig:scaling} shows that there is a scaling law as a functional of the number of training parameters.
Here, we set $N_{\rm MC}^{\rm eff} = 800$.  
Blue square corresponds to the estimated MSE for the linear model and red circles indicate MSE for $L=1,2,\cdots,6$ from the left.
We fit only points for $L\geq2$ to guide eyes.

We remark that in the SLMC, new training data is always generated so that the dataset size is increasing during the SLMC. 
This means that large data requirement for training of Attention layers can be satisfied by self-training. 
Moreover, we can systematically increase the number of trainable parameters in our model with adding new attention layer without loosing performance. 

We discuss the origin of the scaling law that we found. 
We note that the origin of the scaling law in large language models is not well understood. 
Although we do not know the direct origin of our scaling law, we show that the MSE of the effective model can be improved with increasing the number of the attention layers.  
We introduce the parameter space of the effective model with $L$ attention layers ${\cal P}^{(L)}$. 
The parameter space with different number of the attention layers has the relation expressed as  
\begin{align}
{\cal P}^{(0)} \in {\cal P}^{(1)} \in \cdots \in {\cal P}^{(L)} \in \cdots,
\end{align}
where ${\cal P}^{(0)} $ is the parameter space of the linear model. 
Let us consider the model with layer $L$ which has been optimized. 
In this case, the model loss has reached to the lower bound. 
And we add one more attention block with small random weights to the model. 
The model obtains additional capacity for lowering the loss.
Note that, if the additional self-attention block with weight zero, the additional attention layer behaves as an identity map since it is connected with the residual connection.
Thus, the MSE of the effective model with $L$ attention layers ${\rm MSE}^{(L)}$ has the following relation:
\begin{align}
     {\rm MSE}^{(0)} \ge  {\rm MSE}^{(1)} \ge \cdots \ge {\rm MSE}^{(L)} \ge \cdots.
\end{align}
This relation suggests that we can increase the number of training parameters systematically.

\subsection{Future direction}
As a future work, we need to examine volume $N$ scaling and the improvement of our symmetry equivariant attention approach in SLMC. 
Our attention will also be given to adapting this methodology to diverse models.
We also plan to investigate and improve upon the sub-optimal volume scaling associated with the use of flow-based sampling algorithm in lattice field theory \cite{Del_Debbio_2021,deldebbio2021machine, komijani2023generative}.
These strides could potentially improves acceptance ratio and elevate simulation efficiency across the board.

\section{Summary} \label{sec:summary}
We introduced a new type of Transformer network with an attention layer that has spin-rotational and translational equivariance. 
The effective model with the attention layers is introduced in the double exchange model, where electrons are coupled to classical spins on a lattice. Using the SLMC, we generate new training data during Monte Carlo simulations to train the effective model. 
In the SLMC, the effective model trained at a single temperature can be applied over the whole temperature range and reproduces the antiferromagnetic phase transition. 
Without using the SLMC, we showed that the MC with effective models almost reproduces the physical quantities obtained by the original model. 
This indicates that the effective model constructed using the transformer architecture has sufficient capability to capture the original model. 
We find that the mean squared error (MSE) decreases with increasing the number of training data, following a scaling law similar to that observed in Transformer networks for large language models. 
Our effective model is a natural extension of the linear model and captures the nonlinear behavior of target systems.

\begin{acknowledgments}
The work of A.T. was partially by JSPS  KAKENHI Grant Numbers 20K14479, 22H05112, and 22H05111.
Y.N. was partially supported by JSPS KAKENHI Grant Numbers 22K12052, 22K03539, 22H05111 and 22H05114. 
The calculations were partially performed using the supercomputing system HPE SGI8600 at the Japan Atomic Energy Agency.
This work was supported by MEXT as ``Program for Promoting Researches on the Supercomputer Fugaku'' (Simulation for basic science: approaching the new quantum era; Grant Number JPMXP1020230411).
This work was supported by MEXT as ``Program for Promoting Researches on the Supercomputer Fugaku'' (Search for physics beyond the standard model using large-scale lattice QCD simulation and development of AI technology toward next-generation lattice QCD; Grant Number JPMXP1020230409).
\end{acknowledgments}

\appendix

\section{Attention and Transformer}\label{sec:attention}
Here we briefly review Transformer and Attention block \cite{vaswani2017attention}, which have large model capacity \cite{kaplan2020scaling}. 
Please see \cite{lin2021survey} for detail and recent development.

The attention layer is essential component of the transformer neural networks.
The input consists of queries, keys, and values of dimension $d$. 
In the conventional attention layer, so-called scaled dot-product attention layer, we compute the dot products of the query with keys, divide each by $\sqrt{d}$ and apply the activation function to obtain the weights of the values. According to the Ref.~\cite{vaswani2017attention}, the conventional attention layer is defined as 
\begin{align}
    {\rm Attention}(Q,K,V) = {\rm softmax} \left(\frac{Q K^T}{\sqrt{d}} \right) V.
\end{align}
Here, $Q$,$K$ and $V$ are tensors whose size depend on system. 
The self-attention layer defined as 
\begin{align}
    {\rm SelfAttention}(x) = {\rm Attention}(W^{\rm Q}x,W^{\rm K}x,W^{\rm V}x), 
\end{align}
is used in the Transformer. 
Here, $W^{\rm Q}$, $W^{\rm K}$ and $W^{\rm V}$ are trainable tensors. 

In the first paper \cite{vaswani2017attention},
they develop multi-head attention, which is constructed by output of single-head output attention explained here. 
In this paper, we utilize single-head attention for simplicity.

\section{Self-learning Monte Carlo}\label{sec:slmc}
The physical observable such as the magnetization can be calculated by the MCMC. 
In the Monte Carlo method, we have to generate a spin configuration $\{ {\bm S} \} $ with a probability distribution $W(\{ {\bm S} \} )$. 
By constructing a Markov chain that has the desired distribution as its equilibrium distribution:
\begin{align}
    \{ {\bm S} \}_{1} \rightarrow   \{ {\bm S} \}_{2}  \rightarrow  \cdots\rightarrow  \{ {\bm S} \}_{i} \rightarrow  \cdots ,
\end{align}
we can obtain a sample of the desired distribution by observing the chain after a number of steps. 
We introduce the condition of the detailed balance expressed as 
\begin{align}
W(\{ {\bm S} \} ) P(\{ {\bm S}' \}| \{ {\bm S} \}) &= W(\{ {\bm S}' \} ) P(\{ {\bm S} \}| \{ {\bm S}' \}),
\end{align}
where $P(\{ {\bm S}' \}| \{ {\bm S} \}) $ is the transition probability from another configuration $\{ {\bm S} \}$ to a configuration $\{ {\bm S}' \}$.
In the Metropolis-Hastings approach, the transition probability is separated in two substeps:
\begin{align}
     P(\{ {\bm S}' \}| \{ {\bm S} \}) &= g( 
     \{ {\bm S}' \} |  
     \{ {\bm S} \}) 
     A( \{ {\bm S}' \}, \{ {\bm S} \}),
\end{align}
where the proposal distribution $g( \{ {\bm S} \} '|  \{ {\bm S} \})$ is the conditional probability of proposing a configuration $\{ {\bm S}' \}$ when a configuration $\{ {\bm S} \}$ is given, and the acceptance ratio $A( \{ {\bm S}' \}, \{ {\bm S} \})$ is the probability to accept the proposed configuration $A( \{ {\bm S}' \}, \{ {\bm S}' \})$. 
The Markov chain that has the desired distribution $W(\{ {\bm S} \} ) $ is obtained when the acceptance ratio is given as 
\begin{align}
    A( \{ {\bm S}' \}, \{ {\bm S} \}) = {\rm min} \left( 1, \frac{W(\{ {\bm S}' \} ) }{W(\{ {\bm S} \} ) } \frac{g( \{ {\bm S} \} |  \{ {\bm S}' \})}{g( \{ {\bm S}' \} |  \{ {\bm S} \})} \right). \label{eq:A}
\end{align}

One can design various kinds of the Monte Carlo method based on Eq.~(\ref{eq:A}). 
The most simple update method is so-called local update, where the spin configuration is updated locally. 
A single site is randomly chosen in the current configuration and a new configuration is proposed by changing the orientation of the classical spin. 
Although it is easy to implement the local update in various kinds of systems, the computational cost is usually high because the autocorrelation time is long. 
We note that the hybrid Monte Carlo method is known as the one of the good update methods, which is widely used in lattice QCD.

\subsection{Basic concept of the SLMC}
In MCMC, the proposal probability $g( \{ {\bm S}' \} |  \{ {\bm S} \})$ can be designed to increase the acceptance ratio $A( \{ {\bm S}' \}, \{ {\bm S} \})$. 
If the ratio of the proposal probability $\frac{g( \{ {\bm S} \}  \{ {\bm S}' \})}{g( \{ {\bm S}' \}  \{ {\bm S} \})} = \frac{W(\{ {\bm S} \} ) }{W(\{ {\bm S}' \} ) }$, the new configuration $\{ {\bm S}' \}$ is always accepted because of $A( \{ {\bm S}' \}, \{ {\bm S} \})=1$. 
In the SLMC, we introduce the another Markov chain with the probability $W_{\rm prop}(\{ {\bm S} \})$. 
The detailed balance condition is given as 
\begin{align}
W_{\rm prop}(\{ {\bm S} \} ) P_{\rm prop}(\{ {\bm S}' \}| \{ {\bm S} \}) 
= W_{\rm prop}(\{ {\bm S}' \} ) P_{\rm prop}(\{ {\bm S} \}| \{ {\bm S}' \}).
\end{align}
On the proposal Markov chain, the configuration $\{ {\bm S}' \}$ is obtained by the random walk from $\{ {\bm S} \}$. 
The proposal probability $P_{\rm prop}(\{ {\bm S}' \}| \{ {\bm S} \}) $ can be regarded as the conditional probability $g( \{ {\bm S}' \} |  \{ {\bm S} \})$ on the original Markov chain. 
Thus, the acceptance ratio in the SLMC is given as 
\begin{align}
    A( \{ {\bm S}' \}, \{ {\bm S} \}) = {\rm min} \left( 1, \frac{W(\{ {\bm S}' \} ) }{W(\{ {\bm S} \} ) } 
    \frac{W_{\rm prop}(\{ {\bm S} \} ) }{W_{\rm prop}(\{ {\bm S}' \} ) } 
\right). 
\end{align}
If we can design the proposal Markov chain whose probability is equal to that of the original Markov chain, the proposed spin configuration $\{ {\bm S}' \}$ is always accepted. 

\subsection{Model in previous work}
In the DE model, the long-range correlation is partially considered in the linear effective model \cite{Liu2017-wl,Kohshiro2021-ea}:
\begin{align}
\Ham_{\rm eff}^{\rm Linear} = E_0 - \sum_{\langle i,j \rangle_n} J_n^{\rm eff} {\bm S}_i \cdot {\bm S}_j. \label{eq:linear}
\end{align}
With the use of the SLMC, we can fit the effective coupling $J_n^{\rm eff}$ far from the weak coupling regime and the functional form of $J_n^{\rm eff}$ is similar to the RKKY interaction \cite{Liu2017-wl,Kohshiro2021-ea}. 
However, the model capacity is poor because this is the linear model with respect to the two-body terms ${\bm S}_i \cdot {\bm S}_j$. 
Therefore, with the use of the Transformer architecture, we include the long-range correlation in the effective model.

\bibliography{apssamp}
\clearpage

\end{document}